\documentclass[12pt]{article}
\usepackage{epsfig}
\usepackage{amsfonts}
\newcommand{\pip}    {\mbox{$\pi^{+}$}}
\newcommand{\piz}    {\mbox{$\pi^{0}$}}

\begin{document}
\begin{sloppypar}

\begin{center}
\begin{Large}
{\bf Study of the reaction} $\mathbf{pp \to pp\piz}$ {\bf within 10 
MeV above the threshold}

\end{Large}
\vspace*{5mm}

COSY-TOF collaboration

S.~AbdEl Samad$^e$, R.~Bilger$^d$, A.~B\"ohm$^b$, K.-Th.~Brinkmann$^b$,
H.~Clement$^d$, S.~Dshemuchadse$^f$, W.~Eyrich$^c$, D.~Filges$^e$, 
H.~Freiesleben$^b$, M.~Fritsch$^c$, R.~Geyer$^e$, D.~Hesselbarth$^e$,
B.~Jakob$^b$, L.~Karsch$^b$, K.~Kilian$^e$, H.~Koch$^a$,
J.~Kress$^d$, E.~Kuhlmann$^{b\#}$ , S.~Marwinski$^e$, P.~Michel$^f$, 
K.~M\"oller$^f$, H.P.~Morsch$^e$, L.~Naumann$^f$, M.~Richter$^b$, 
E.~Roderburg$^e$, M.~Rogge$^e$, A.~Schamlott$^f$, M.~Schmitz$^e$, 
P.~Sch\"onmeier$^b$, W.~Schroeder$^c$, M.~Schulte-Wissermann$^b$, 
M.~Steinke$^a$, F.~Stinzing$^c$, G.Y.~Sun$^b$, G.J.~Wagner$^d$, 
M.~Wagner$^c$, A.~Wilms$^a$, S. Wirth$^c$ 

\end{center}

%\vspace*{3mm}

\begin{footnotesize}

$^a$ Institut f\"ur Experimentalphysik, Ruhr-Universit\"at Bochum, 
D-44780 Bochum, Germany\\
$^b$ Institut f\"ur Kern- und Teilchenphysik, Technische
Universit\"at Dresden, D-01062 Dresden, Germany\\
$^c$ Physikalisches Institut, Universit\"at Erlangen, D-91058
Erlangen, Germany\\
$^d$ Physikalisches Institut, Universit\"at T\"ubingen, D-72076
T\"ubingen, Germany\\
$^e$ Institut f\"ur Kernphysik, Forschungszentrum J\"ulich, 
D-52425 J\"ulich, Germany\\
$^f$ Institut f\"ur Kern- und Hadronenphysik, Forschungszentrum 
Rossendorf, D-01314 Dresden, Germany

\end{footnotesize}

\begin{center}
April 25, 2003\\

\vspace*{5mm}

{\bf Abstract}
\end{center}

\begin{small}

\noindent
%\vspace{3mm}

Kinematically complete measurements of the $pp\to pp\piz$ reaction were
performed for beam energies in the range $292 - 298\,$MeV. We detected
both protons in coincidence by using the large acceptance COSY-TOF
spectrometer and an external proton beam at COSY-J\"ulich; thus we
measured total and differential cross sections and energy distributions.
A strong enhancement was observed in the Dalitz plots due to the final 
state interaction between the two outgoing protons; the data are well 
reproduced by Monte Carlo simulations using a standard scattering 
length of $a_0$=-7.83 fm and effective range of $r_0$=2.8 fm. Our 
measured total cross sections are roughly $50\,\%$ larger than those of 
recent internal target experiments at IUCF and CELSIUS. This 
discrepancy may be due to the final state interaction pushing some 
events into the very small-angle region which is almost inaccessible to
an internal target experiment. In the angular distributions we found
only very slight deviations from isotropy. The analysis of our
momentum distributions yielded a small fraction of $5\pm 5 \%$ for Ps 
or Pp wave contributions.

\vspace{3mm}

$\#$ corresponding author, e.kuhlmann@fz-juelich.de
\end{small}

\newpage

\begin{large}

{\bf I. Introduction}

\end{large}

The last decade saw a renaissance of near-threshold pion producing
reaction studies in nucleon-nucleon (NN) collisions. With the advent
of medium-energy accelerators in Bloomington/USA (IUCF), Uppsala/Sweden
(CELSIUS), and J\"ulich/Germany (COSY) a wealth of high precision data
on total and differential cross sections as well as polarization
observables was measured for the reactions $pp \to pp\piz$ 
\mbox{$\lbrack$ 1-6 $\rbrack$}, $pp \to pn\pip$ \mbox{$\lbrack$ 7-9 
$\rbrack$} and $pp \to d\pip$ \mbox{$\lbrack$ 10-11 $\rbrack$}, which 
in turn induced a flurry of theoretical activity \mbox{$\lbrack$ 12-18 
$\rbrack$}. So far, good agreement is found in the theoretical
description of the reactions $pp \to d\pip$ and $pp \to pn\pip$, larger
discrepancies are observed for the $pp \to pp\piz$ reaction. Soon after
publication of the near-threshold data on neutral pion production it
was recognized that the $pp\piz$ reaction near threshold is sensitive
to short-range mechanisms in the NN-system. Since the main pion
exchange term, which dominates the charged pion producing channels, is
isospin-forbidden, Lee and Riska proposed to counter this shortfall by 
the additional consideration of pair diagrams with an exchanged
heavy meson ($\sigma, \omega$) \cite{lee}. Likewise the role
of pion-rescattering only then was treated with more concern, but also 
in a somewhat controversial manner, since field theoretical models and
chiral perturbation theories found different relative signs in the
$\pi$-exchange amplitude with respect to the direct (Born) term 
\cite{park}, \cite{cohen}, \cite{han1}. Attempts have been pursued
to additionally include $\Delta (1232)$ isobar and $S_{11}$ and 
$D_{13}$ nucleon resonance excitations \mbox{$\lbrack$ 15-17 $\rbrack$}.
To date, calculations in the framework of the J\"ulich
meson exchange model \cite{han1}, \cite{han2}, which incorporate all
the basic terms like heavy meson exchange (HME), delta resonance
and off shell effects as well as realistic final-state interactions,
yield the best results; with the exception of the strength parameter
of the HME term the model needs no adjustable parameters. 

Close to threshold only a few partial waves have to be considered with 
Ss, Sp, Ps and Pp being the leading ones, whereas the role of $l=2$ 
contributions (Sd and Ds) is still under debate. Here the Rosenfeld
notation $L_{p}l_q$ has been used \cite{rose} with $L_p$ being 
the orbital angular momentum of the NN-pair, $l_q$ that of the 
pion with respect to this pair. In the $pp\piz$ reaction the partial 
wave Sp which apart from Ss governs the isoscalar reaction channels 
$d\pip$ and $pn\pip$, is forbidden by conservation laws. As a result 
purely isotropic angular distributions are found for the $pp\piz$ 
reaction in the region up to 10 MeV above threshold \cite{bond}. Also, 
sizeable analyzing powers are only observed for this reaction for excess
energies Q higher than 16 MeV \cite{hom2} which follows from the 
fact that interference effects are only possible between amplitudes (Ps,
Pp) on the one hand or (Ss,Sd,Ds) on the other. Except for Ss, however, 
the magnitude of all these amplitudes is negligibly small up to 
Q =16 MeV. For comparison, angular distributions as well as 
analyzing powers extracted from a study of the $\pip$-producing 
reactions show clear evidence for the presence of Sp, and hence $l=1$ 
contributions at energies as low as 2 MeV above threshold 
\mbox{$\lbrack$ 10-11 $\rbrack$}.

Measurements very close to threshold, where all escaping particles
are confined to a narrow cone around the beam axis, require a perfectly
matched detector system which has to cover as much of the 
available phase space as possible in order to keep
necessary acceptance corrections to a minimum. In J\"ulich, 
the COSY-TOF spectrometer,
a scintillator hodoscope of cylindrical shape with full 
$\phi$-symmetry fulfills these requirements to a very large extent.
Charged particles are detected at polar angles as close as one degree 
to the beam. In the past we have used this spectrometer 
in meson- and hyperon-production studies and pp bremsstrahlung 
experiments. Parallel to the analysis of the bremsstrahlung reaction 
\cite{herm}, data obtained simultaneously for $pp \to pp\piz$
were analysed as a cross check.  We systematically found larger 
cross sections than given in \cite{hom1}, \cite{bond}. Since the results
we deduced from the data for the concurring charged $\pi$-producing 
reactions agreed nicely with the published values \cite{hard}, 
\cite{gem2}, \cite{heim}, whereas the cross sections for 
$\piz$-production remained high, we decided to perform a second 
experiment to clarify that matter.

In this paper we will present data from a kinematically complete
measurement of the reaction $pp \to pp\piz$ for three energies in the
range 292 - 298 MeV, corresponding to beam momenta of 
796-805 MeV/c. Total cross sections as well as angular and 
energy distributions will be shown and compared to Monte Carlo 
simulations which are based on phase space distributed events that are 
modified by the final-state interaction (FSI). Results of the concurring
reactions $pp\to d\pip$ and $pp\to pn\pip$ which were obtained in
parallel will only be given to underline the performance of the
spectrometer and to prove the consistency of the analysis. Apart from 
the excess energy Q we will also use the dimensionless parameter 
$\eta$ for labelling energies above threshold defined as $\eta 
=q_{max}/ m_{\pi}$ with $q$ being the CM-momentum of the pion and 
$m_{\pi}$ its mass. With $p$ we denote the CM-momentum of one proton in 
the two-proton subsystem. The angles $\Theta_q$ and $\Theta_p$ (in the 
CM-system) are taken with respect to the beam. The momenta $q$ and $p$ 
are linked by energy conservation via $\sqrt{s}=\sqrt{m^2_\pi +q^2} +
\sqrt{M^2_{pp} +q^2}$ and $M^2_{pp}=4(m^2_p + p^2)$ where $s$ is the 
square of the total CM energy
and $M_{pp}$ the invariant mass of the two protons.

\newpage
\begin{large}

{\bf II. Experimental Procedure}

\end{large}

{\bf A. Apparatus}

The experiment was carried out with the time-of-flight spectrometer
COSY-TOF \cite{dahm} \cite{bohm} set up at an external beamline of the 
2.5 GeV proton synchrotron COSY at J\"ulich/Germany. The goal was to
study simultaneously and with the same spectrometer all five reaction 
channels which are open in pp-collisions at beam energies near 
300 MeV. Besides pp-elastic scattering, which is by far the 
strongest, and the very weak pp bremsstrahlung, our interest mainly 
aimed at the $\pi$-producing reactions $pp \to d\pip$, $pp\to pp\piz$, 
and $pp\to pn\pip$, each one with cross sections rising fast with 
excess energy. 

\begin{figure}[htb]
\vspace*{0.6cm}
\begin{center}
\epsfig{file=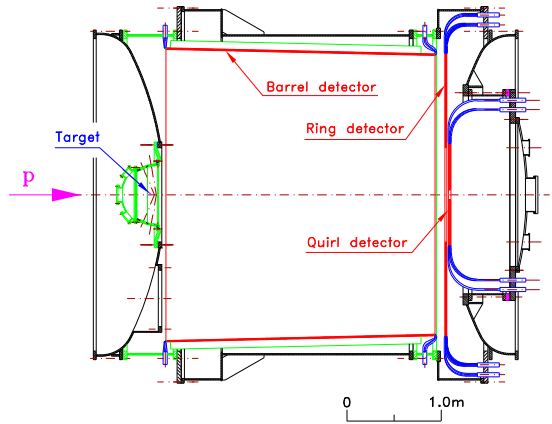,scale=0.5}
\end{center}
\vspace*{-0.3cm}
\caption{\it Sketch of the COSY-TOF spectrometer.}
\end{figure}

A schematic sketch of the spectrometer showing the locations of the 
main components, the barrel (B), ring (R), and quirl (Q) detector, 
as well as the target region is given in fig.~1. To prevent the
reaction products from undergoing secondary reactions on their way
from the target to the different detector modules of the spectrometer
it is housed within a 3.3 m $\oslash\,\times$ 4 m steel tank which can 
be evacuated to a pressure of 0.2 Pa. Protons were 
extracted out of the COSY ring in a slow extraction mode allowing spill 
lengths of up to 10 min. Their intensity typically was of the order 
of several $10^6$/s. The beam was focussed onto the liquid hydrogen 
target with dimensions 6 mm $\oslash\,\times$ 4 mm \cite{hassan}. Its 
front and rear ends were closed by 0.9 $\mu$m thin hostaphan foils. A 
set of scintillator veto detectors with central holes of various sizes 
were located 260, 51 and 3.5 cm upstream of the target. They helped 
to define the beam spot in the center of the target to an area with 
diameter d=3.0 mm. Charged particles emitted from the target into the 
forward hemisphere had to transverse a 0.5 mm thin plastic scintillator 
device \cite{michel} which served as start detector. This detector 
shown in fig.~2 consists of two concentric rings made out of 16 
trapezoidally shaped scintillators each, which were placed 30 mm (Ring
B) and 50 mm (Ring A) behind the target. The inner detector had a 
central hole of diameter d$_i$=3.3 mm and extended out for 26 mm, the
outer one had dimensions d$_i$=16 mm and d$_a$=130 mm. The minimum 
angle for observable ejectiles was $1.9^\circ$ if they originated from 
the center of the target, but smaller values far less than $1.0^\circ$ 
were possible if the initial pp-reaction
took place off the beam axis. The elements of the outer ring were 
arranged in such a way as to show a tiny overlap along their sides thus 
guaranteeing full $\phi$-coverage whereas the ones of the inner ring
were only allowed to touch each other thereby leaving narrow gaps of
order 50 $\mu$m which caused an overall loss in $\phi$-coverage of 
$3\,\%$ distributed evenly over the full circle.

\begin{figure}[htb]
\vspace*{1.0cm}
\begin{center}
\epsfig{file=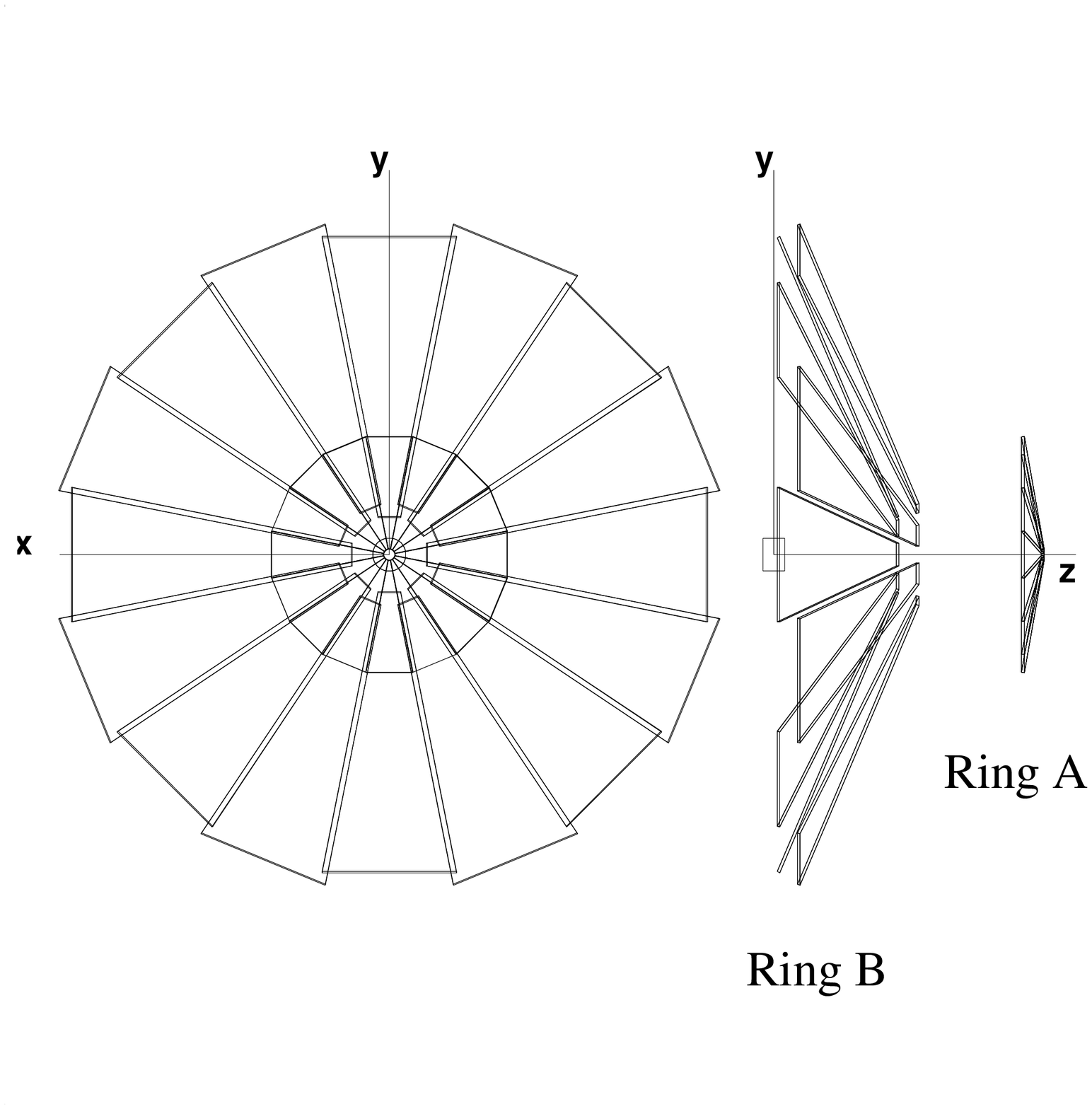,scale=0.5}
\end{center}
\vspace*{-0.3cm}
\caption{\it Front and side view of the start detector. The target is
placed to the left of ring B.}
\end{figure}

After a flight path of up to 3 m in vacuum, the reaction products hit 
the 3-component (B, R, and Q) scintillator hodoscope. 
The circular, 3-layer quirl and ring detector components of the endcap
were both built according to the same layout. 5 mm thin scintillator 
sheets (BC408) were used as basic material. The first layer in the 
quirl is made from 48 wedge-like scintillators, followed by two layers
with 24 left and 24 right wound elements, each element cut along an
Archimedian spiral \cite{dahm}. In central projection a net-like 
pattern is formed consisting of more than 1000 triangular pixels. A 
central hole of radius r=4.2 cm was left for the beam, the outer 
radius of the quirl is 58 cm. The three layers in the ring were made 
from 96 wedges and $2\times 48$ left and right wound Archimedian 
spirals, the inner (outer) radius is 56.8 (154) cm, respectively. 
The barrel consists of 96 scintillator bars, mounted to the inside of 
the tank. Each bar of 2.85 m length has a cross section with 
measures thickness $\times$ width =15 mm $\times$ 96 mm and is read 
out from both ends \cite{bohm}.

In the course of the experiment which went on for almost two years
two slightly different set-ups were employed. The one described above
with barrel, ring and quirl was used at T=293.5 MeV. Azimuthal 
coverage was complete, the acceptance in polar angle $\theta$ ranged 
from $1.5^\circ - 78^\circ$. In an earlier experiment performed at 
T=292.2 and 298.1 MeV the ring component was still under 
construction, leaving a gap in polar angle $\theta$ between $10.4^\circ 
- 26^\circ$. For this run a modular neutron detector \cite{leo} was 
additionally set up downstream around the beam line and outside of the 
tank. It covered approximately the same area as the quirl detector and 
was employed in the study of the concurring $pp\to pn\pip$ reaction.
As a first level trigger usually one, in some cases two charged hits 
were required in the start detector, at least two in any of the stop 
detectors B, R, Q, with 2B, 2R, 2Q and 2(BRQ) denoting separately 
handled trigger patterns. For each of these patterns slightly different 
deadtimes were observed and the necessary corrections had to be applied 
accordingly. 

\vspace*{0.5cm}

{\bf B. Performance Checks and Luminosity Monitor}

Simultaneously with the $\pi$-producing reactions, elastic scattering 
events, which were detected either as barrel-barrel (BB) or as 
barrel-ring (BR) coincidences, were written on tape. 
In addition to serving as a tool for calibrating the 
whole detector system, this reaction was used as a luminosity monitor
through comparison of experimentally deduced elastic scattering events
with data taken from the SAID database \cite{said} (see fig.~3).
The cross section around T = 300 MeV is known with an 
error of less than $5\,\%$. Any cross section of interest determined 
relative to the one for $pp_{elast}$ will thus be measured with 
comparable accuracy, since all uncertainties resulting from
instabilities in beam intensity and target thickness cancel. 

\begin{figure}[htb]
\begin{center}
\epsfig{file=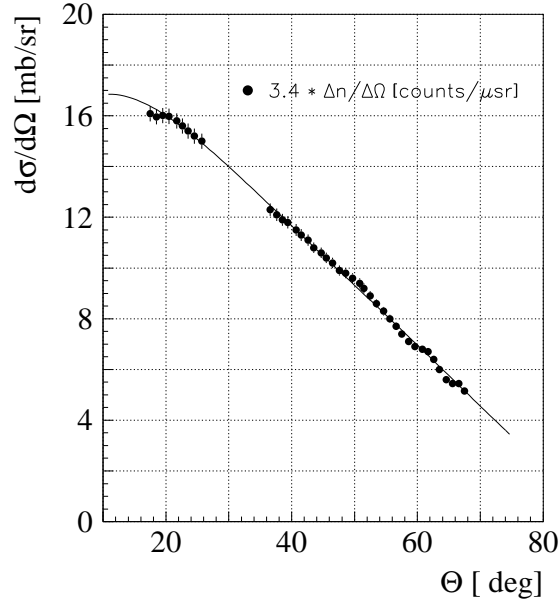,scale=0.45}
\end{center}
\vspace*{-1.5cm}
\caption{\it Measured angular distribution of elastic pp
scattering events (solid dots) compared to the SAID solution (full 
line, see text).}
\end{figure}

We obtained complementary information on the detector performance 
from a careful study of the $pp\to d\pip$-reaction. Its cross section 
at 295 MeV is close to 45 $\mu$b \cite{gem2}. While the deuterons with 
a maximum polar angle at $3^\circ$ will only reach the quirl, the pions 
with $\theta^{max} = 32^\circ$ can be observed in all of the three 
stop-detector components. The unique kinematics of this 2-particle 
reaction allowed the extraction of rather clean and almost background 
free spectra as e.g., invariant $d$- and $\pip$-mass distributions. 
In the present case this reaction was used for an accurate 
determination of the beam energy $T_p$, which is important in view of 
the strong $T_p$ dependence of the cross section as well as available 
phase space coverage, which in turn has a strong impact on the 
acceptance correction. In fig.~4 the transverse 
$\pi$-momentum $p_t$ is plotted versus the longitudinal $\pi$-momentum 
$p_l$. The boxes represent the experimental data, their sizes being 
proportional to the respective number of counts. Concentration of yield 
is observed along an elliptical curve which aligns nicely with the one 
resulting from two-body kinematics for a beam energy of T$_p$ =293.5 
MeV (dashed curve). The two solid lines calculated for beam energies 
which are 0.6 MeV higher and lower clearly miss the main body of the 
data from which we deduce an error in beam energy of order 0.3 MeV. 

\begin{figure}[htb]
\vspace*{-0.5cm}
\begin{center}
\epsfig{file=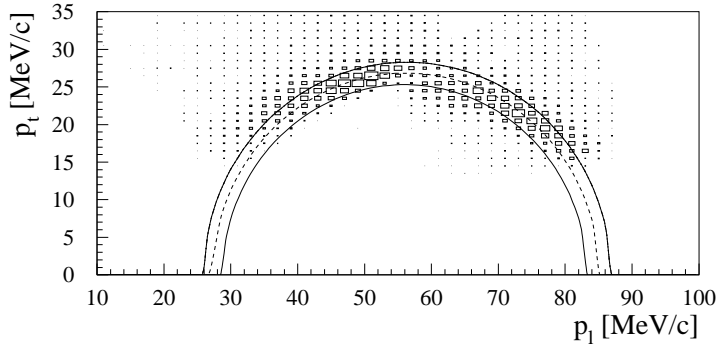,scale=0.5}
\end{center}
\vspace*{-6.0cm}
\caption{\it Transverse vs. longitudinal $\pi$-momentum in 
comparison with expectations from two-body kinematics for three
different beam energies (see text for details).}
\end{figure}

The modulation in yield along the elliptical curve and the 
absence of yield for $p_t$-values below 17 MeV/c is well understood and 
can be explained as stemming from particular technical details in 
detector construction as well as the rather low pion momenta: pions 
emitted at angles larger than $26^\circ$ will be observed in the 
one-layer barrel section; these are the events with $p_l$ up to 52 
MeV/c. Those with very low momenta, however, will not be able to reach 
it. The ring section covers the range $10.4^{\circ}-26^\circ$; here, 
however, a pion has to reach at least the second layer in order to fix
its point of impact, hence a decrease in yield is observed for pion 
angles approaching the outer range of the ring. Very forward pions with 
low $p_t$-values were not observed, since the coincident deuterons 
escaped through the central hole in the start detector. 

\vspace*{0.5cm}

\begin{large}

{\bf III. Data Analysis and Calibration}

\end{large}

{\bf A. Principle of Measurement}

The photomultiplier signals of all detector elements were used for 
energy-loss as well as time-of-flight measurements. To this end, the 
analog signals were digitized in Fastbus-QDC modules, the timing signals
were derived in leading-edge type discriminators and used in Fastbus 
TDCs. Both measurements yield the particle velocity $\beta$. From the 
particle's point of impact the flight direction was obtained from which 
two additional observables, $\theta$ and $\phi$, can be derived. 
In the endcap with its 3-layer quirl and ring components the flight 
direction is deduced by combinatorical means \cite{dahm}. For the 
one-layer barrel elements with two-sided readout the point of impact 
is found from the difference in TDC-values of both ends \cite{bohm}, 
whereas the TDC-sum is a measure of the flight time. The spatial 
resolution as given from the pixel size in the endcap is 
$\Delta \theta = 0.5^\circ, \Delta \phi = 2^\circ$; roughly the same 
values hold for the barrel elements. Typical values obtained for the 
time resolution of the start-stop system were of the order of 0.5 ns 
(FWHM) which corresponds to about $2.5\,\%$ of the average flight time 
of a $\beta = 0.5$ particle. 

The momentum 4-vector of each detected particle can then be deduced
from the measured observables $\beta$, $\theta$, and $\phi$ by 
applying an additional mass hypothesis. This of course has to be
checked in comparison with results obtained from Monte Carlo 
simulations. In case of a reaction with one neutral particle like $pp 
\to pp\piz$ the 4-vector of that particle is deduced by employing 
energy and momentum conservation and calculating its mass.

%\vspace*{0.5cm}
\newpage

{\bf B. Calibration}

In order to determine $\theta$, $\phi$ and $\beta$ for each particle,
several steps in calibrating the detector had to be performed 
\cite{jakob}. After a simple pedestal subtraction in the QDC spectra, a 
walk correction had to be applied for each TDC entry. 
Depending on the pulse height of the scintillator signal, time shifts 
of up to 3 ns were observed. To correct for these shifts a 
five-parameter fit function was found which relates the observed walk
with the pulse height given by the corrected QDC value. The complete 
parameter list for up to 600 scintillator channels was determined 
before each experiment. This was routinely done by use of a laser-based 
calibration system where UV-light from a $N_2$-laser is fed via 
quartz fibers into each of the individual scintillator channels.
Through use of a set of filters the amount of light going into the 
various elements can be varied such as to cover the complete 
dynamic range of the multipliers. 

A check on the differential nonlinearity in the TDC modules revealed no 
deviations from the precision given by the manufacturer ($<0.1\,\%$ of 
full scale), in some cases, however, a channel width of 
90 ps/channel was observed instead of the preset value of 
100 ps/channel. Due to different cable lengths and varying transit 
times in the photomultiplier tubes and electronic modules, variations 
in the signal arrival time arise which are specified as TDC-offsets. 
These offsets were determined in an iterative manner by first comparing 
overlapping, but otherwise identical channels as e.g., neighboring 
elements in the outer start detector (sect.~II.A) or overlap regions 
from left- and right-wound elements in the endcap. Two-particle 
reactions like $pp_{elast}$ and $d\pip$ with their unique kinematics 
then were used for the adjustment of all remaining detector components. 
As a byproduct the light velocities within the differently shaped 
scintillator elements were deduced. Whereas in bulk material it is 
given by $c/n$ ($n$: index of refraction), 
smaller values were found for the rather flat, but long elements due to 
an effective lengthening of the lightpath. For the straight quirl 
elements of wedgelike shape and a thickness of 5 mm we found 
18.0 cm/ns, for the corresponding ones of the ring 15.8 cm/ns. 
A value of 16.1 cm/ns was observed for the 3 m barrel elements.
The calibration of the neutron detector 
with its 10 cm thick bars was performed as outlined in 
\cite{leo}. Special care was applied in the determination of the 
energy-dependent efficiency which, for neutron energies larger
than 60 MeV, was found to be rather constant around $13.5\,\%$.

\newpage
%\vspace*{0.5cm}

{\bf C. Monte Carlo simulation}

Throughout the analysis each step was compared to the results deduced 
from the Monte Carlo simulation. The program package written in 
$C^{++}$ was developed in Bochum \cite{brand}, \cite{ziel} with
the purpose of simulating the spectrometer response, thereby giving 
insight into the overall performance, and to determine the detector
acceptance. At its core is the CERNLIB random event generator GENBOD 
\cite{cern1}, which generates a preset number of N-body events for a 
given reaction specified by N, the number of ejectiles, their masses and
the 4-momentum of the incoming beam. It returns momentum 4-vectors for 
each particle in the overall center of mass system, and weight factors 
$w_{e}$ based on the phase space density of the reaction. These weight 
factors have to be modified, if angular momentum or final-state 
interaction effects are believed to play a substantial role. The Monte 
Carlo code then boosts this event into the laboratory system and each 
particle is tracked through the complete setup. Nuclear interactions
are treated according to the INC code \cite{cloth} which has its roots
in the Bertini intranuclear cascade model of the late sixties 
\cite{bert}. This model assumes the nucleons within their respective
nuclei to behave like a fermigas and the interactions between the
incident particle and the nucleus is simulated as the output of an
intranuclear particle shower with nucleon-nucleon and pion-nucleon
reactions. Elastic and inelastic NN and $\pi$N cross sections are
supplied via extensive data tables. Great care was put into 
modelling all of the detector components in their particular shape as 
close as possible to reality, including the left- and right-wound 
spirals in the endcap and the trapezoidal elements of the slightly 
conical barrel detector. Options were built into the program to study 
in detail the influence of the extended target and the transverse beam 
spread or the variation in acceptance due to the narrow gaps between 
neighboring scintillator elements or the size of the central hole in 
the inner start detector. 

In the course of the simulation, interactions of the particles on their
way through the various active or passive detector parts can be switched
on or off, their energy loss and small-angle scattering can be 
investigated separately. Effective low-energy thresholds can thus be 
studied in detail as eg., in case of the endcap, where a hit in at least
two layers is needed for a successful pixel reconstruction. Also 
included in the Monte Carlo package is the possibility of particle 
decay. In the present case it was found that a sizeable fraction of the
$\pip$-mesons with low momentum had decayed before the detector was 
reached and a decay myon was detected instead. 

%\vspace*{0.5cm}
\newpage

\begin{large}

{\bf IV. Results}

\end{large}

{\bf A. Total cross sections}

By use of energy and momentum conservation, the $pp\piz$ events were 
identified by calculating the missing mass $m_x$ of the unobserved
neutral pion according to 

\begin{center}

$m_x^2 = (\mathbb{P}_b+\mathbb{P}_t-\mathbb{P}_1-\mathbb{P}_2)^2$,

\end{center}

where $\mathbb{P}_b (\mathbb{P}_t)$ denotes the beam (target)
momentum 4-vector, respectively, and $\mathbb{P}_i$ the one of the
i-th proton. The high energy part of the distribution obtained at T=
293.5 MeV is given in fig.~5 showing a clear signal at the expected
location $m_x=m_\pi$= 0.135 GeV/c$^2$.

\begin{figure}[htb]
\vspace*{-0.5cm}
\begin{center}
\epsfig{file=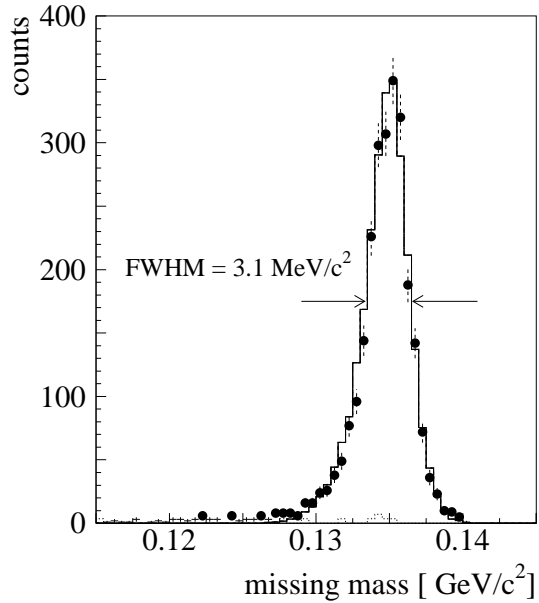,scale=0.45}
\end{center}
\vspace*{-0.7cm}
\caption{\it Missing-mass spectrum of reconstructed $pp\piz$-events
(dots with error bars) in comparison with the Monte Carlo simulation
(histogram). The dashed histogram shows the normalized spectrum 
obtained with an empty target.}
\end{figure}

The only cuts applied in the analysis were upper limits on the 
$\beta$-values of the protons in order to eliminate events where a much 
faster pion was wrongly interpreted as a proton. Background due to 
reactions on the target foils or condensates thereon is negligible as 
can be seen from the spectrum obtained from an empty target run and 
shown by the dashed histogram in fig.~5. Also displayed is the result 
of our Monte Carlo simulation (solid histogram) which is in excellent 
agreement with the data. The angular distribution of the inelastically
scattered protons is shown in fig.~6 (solid dots with error bars)
together with results of our Monte Carlo simulations. A shift towards 
smaller proton angles and away from purely phase space distributed 
events (dashed histogram) is readily seen. When incorporating final 
state interaction effects into the Monte
Carlo code, however, and using standard values for  effective range
$r_0$ and scattering length a, the simulated curve nicely reproduces 
the data as shown by the solid histogram. For more details on our
treating of the final state interaction see sect.~IV.B.

\begin{figure}[htb]
\vspace*{-0.8cm}
\begin{center}
\epsfig{file=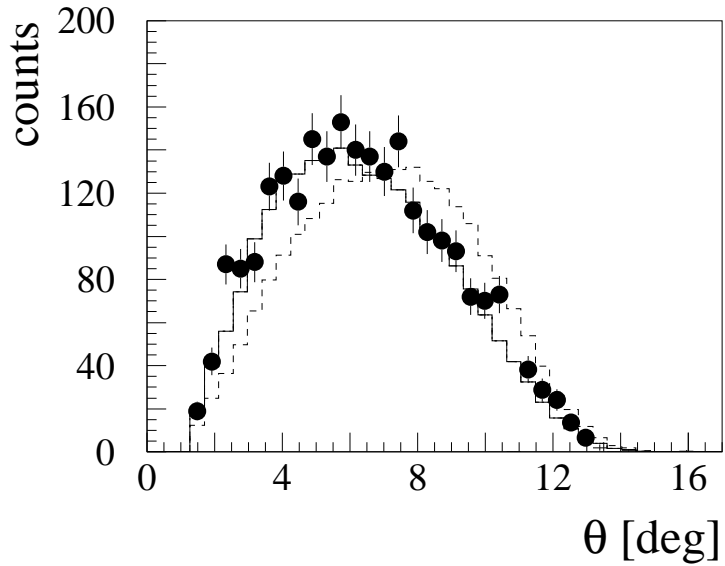,scale=0.85}
\end{center}
\vspace*{-8.5cm}
\caption{\it Proton angular distribution obtained at T=293.5 MeV in
the laboratory system. The solid (dashed) histogram shows Monte Carlo
generated results, where final state interaction effects have been
included (omitted).} 

\end{figure}

In a next step total cross sections for the reaction $pp\to pp\piz$
were extracted from the data by summing up the events for m$_x\geq$ 
0.13\,GeV/c$^2$ (see fig.~5). By making minute corrections for falsely 
interpreted $pp\to pn\pip$-events of order $3\,\%$ from Monte Carlo 
considerations, one finds the number of $pp\piz$-events as given in 
column 3 of table 1. Column 4 lists the acceptances as derived from our 
Monte Carlo simulations. The main deviation from complete acceptance 
($acc\equiv 1$) results from geometrical properties as e.g., the central
hole in the inner start detector (mainly for the low-energy 
measurements) and the missing ring component (see sect.~II.A) for the 
measurement at T=298.1 MeV. Nevertheless, in all cases acceptance 
values of order $50\,\%$ or higher were found. The error in acc was 
estimated by varying in the Monte Carlo code such values as the diameter
of the central hole of the start detector (by some $10\,\%$) or the 
beam energy (by 0.2 MeV). An overall error of $\Delta acc = 3.5\%$ was 
found. The integrated luminosities as given in column 5 of table 1 were 
derived from a comparison with the results obtained for pp-elastic 
scattering in the angular range $32^\circ - 55^\circ$, a range that was 
covered by the barrel alone. For the measurement at T=293.5 MeV, 
where the ring was available as well, the quoted value not only is 
corroborated by the results found for the extended range $18^\circ - 
70^\circ$, but also from a comparison with results for $pp\to d\pip$. 
For this reaction a total cross section of $\sigma = 42 \pm 5\,\mu$b 
was found in very good agreement with an interpolated value of 44 
$\mu$b deduced from \cite{gem3}. For the $pp\to pn\pip$ reaction at 
T=298.1 MeV we determined a cross section of $\sigma_{pn\pip}$=3.5$\pm 
0.4\,\mu b$ which agrees nicely with the interpolated value of 3.1$\pm 
0.4\,\mu b$ extracted from \cite{hard}, \cite{flam}.

%\vspace{0,5cm}
\begin{table}

\caption {\it Total $pp\to pp\piz$ cross sections derived at three beam 
energies. The column labelled $acc$ lists the detector acceptance as 
determined from our Monte Carlo simulations based on FSI modulated 
phase space distributed events.}

\begin{center}

\begin{tabular}{cccccc}

\hline
& & & & & \\
$T\,(MeV)$ &$\eta$ & $N_\pi$ & acc & $\int Ldt\,(nb^{-1})$ & 
$\sigma_{tot}\,(\mu b)$ \\ 
& & & & & \\
\hline
& & & & & \\
292.2 $\pm$ 0.3& 0.29 & 1586 & 0.55 & 1.25 & 2.31 $\pm$ 0.06 $\pm$ 0.23\\
293.5 $\pm$ 0.3& 0.30 & 2524 & 0.64 & 1.41 & 2.80 $\pm$ 0.06 $\pm$ 0.21\\
298.1 $\pm$ 0.2& 0.35 & 4426 & 0.49 & 2.43 & 3.72 $\pm$ 0.06 $\pm$ 0.28\\
& & & & & \\
\hline

\end{tabular}
\end{center}
\end{table}

Total cross sections for the reaction $pp \to pp\piz$ are given in the
last column of table 1 together with their statistical and systematical
errors. The latter are composed of $5\,\%$ from the pp cross section, 
$3.5\,\%$ from the acceptance correction, $1.5\,\%$ from the uncertainty
in misinterpreted background events and $2\,\%$ in the luminosity 
determination.

%\vspace{0,5cm}
\newpage

{\bf B. Dalitz Plots and Energy Distribution}

As mentioned before, a crucial point in the determination of the total 
cross section is the extrapolation of the measured data into 
the full region allowed by phase space. A thorough understanding of the 
detector response is essential, as is the knowledge of any deviation in 
the original (physical) data from those given by simple phase space, 
i.e.~s-wave distributed 

\vspace{-0,5cm}

\begin{figure}[htb]
\vspace*{0.5cm}
\begin{center}
\epsfig{file=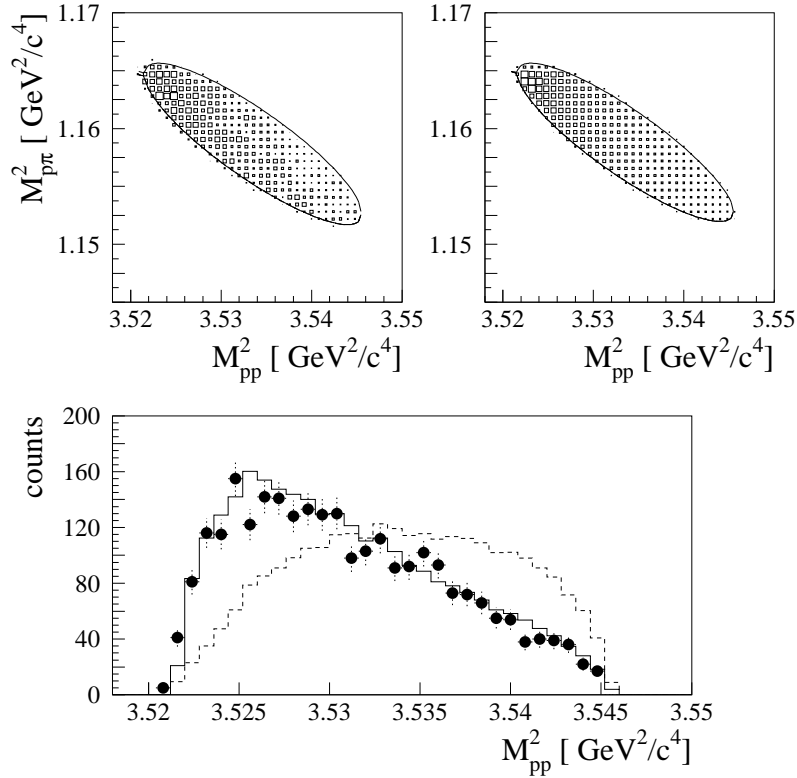,scale=0.55}
\end{center}
\vspace*{-0.7cm}
\caption{\it top: Dalitz plots of experimental (left) and simulated
(right) data with kinematical limits given by the solid lines. The 
enhancement due to FSI is seen in the upper left corner of the event 
distribution. bottom: projection of the Dalitz plots onto the 
$M^2_{pp}$-axis. Experimental data are given by solid dots with 
statistical error bars, the solid histogram shows the simulated 
distribution including FSI effects. The dashed histogram represents the 
simulated distribution for pure phase space.}

\end{figure}

events. A convenient way to compare in a comprehensive manner 
experimental and simulated data from a 3-particle reaction is to set 
up a Dalitz plot, which quite instructively allows the extraction of 
the underlying physics from the event distribution. Purely phase space 
distributed events will result in a uniformly covered Dalitz plot, 
deviations from such a distribution will induce some modulations.
The Dalitz plot as found from the present (kinematically fitted) data 
at T=293.5 MeV is shown in fig.~7 and compared to the result of our 
Monte Carlo simulation (top row left and right, respectively). As a
consequence of the strong final state interaction in the pp system 
a systematic enhancement in the top left corner is observed in the data.
The simulated distribution was obtained by incorporating the FSI 
formalism as developed in \cite{watson}, \cite{migdal} and later refined
by Morton et al. \cite{morton}. In short, one calculates additional 
weight factors $w_{fsi}$ which are multiplied with the 
event weight $w_{e}$ given by the CERNLIB subroutine GENBOD (see also 
sect.~III.C). These factors are given as 

\begin{center}
$w_{fsi} = C^2 \cdot
\lbrack C^4 \cdot T^{CM}_{pp} + \frac{(\hbar c)^2}{m_p c^2} \left( 
\frac{m_p c^2} {2(\hbar c)^2}r_0 \cdot T^{CM}_{pp} -\frac{1}{a_0} 
-\sqrt{2} \frac{\alpha m_p c^2} {\hbar c} \cdot h(\gamma _p)
\right) ^2 \rbrack ^{-1}$,
\end{center}

where $T^{CM}_{pp}$ denotes the pp center of mass kinetic energy 
$T^{CM}_{pp} = M_{pp} -2m_p$ and $C^2$ the Coulomb
penetration factor  

\begin{center}
$C^2 = \frac{2\pi \cdot \gamma_p}{e^{2\pi \gamma_p} - 1}$
\end{center}

with $\gamma_p = \frac{\alpha \cdot \mu_{pp} \cdot c}{p_{pp}}$.
Here $\alpha$ is the fine structure constant, $p_{pp} = \sqrt{2
\mu_{pp} T^{CM}_{pp}}$ and $\mu_{pp}$ is the reduced mass of the 
pp-system. The term with $h(\gamma _p) = \gamma _p^2\cdot 
\sum_{n}^{}\frac{1}{n(n^2+\gamma _p^2)} - 0.5772 - ln \gamma _p$
contributes only little and is often omitted.
From literature we took the standard values $a_0$=-7.83 fm 
and $r_0$=2.8 fm \cite{fsi} as input parameters for the scattering 
length and effective range, respectively, for the two protons in the 
$^1S_0$ state. The smooth rise in yield which readily is seen in the 
simulated data is partially obscured in the experimental data due to the
limited statistical accuracy. However, when inspecting the projection 
of the 2-dimensional Dalitz plot distribution onto the x-axis it is 
obvious that only through inclusion of FSI effects the simulated data 
follow the experimental ones (fig.~7, bottom) whereas a simulation 
based on purely phase space distributed events fails over the whole 
range.

\begin{figure}[htb]

\vspace*{-1.5cm}
\begin{center}
\epsfig{file=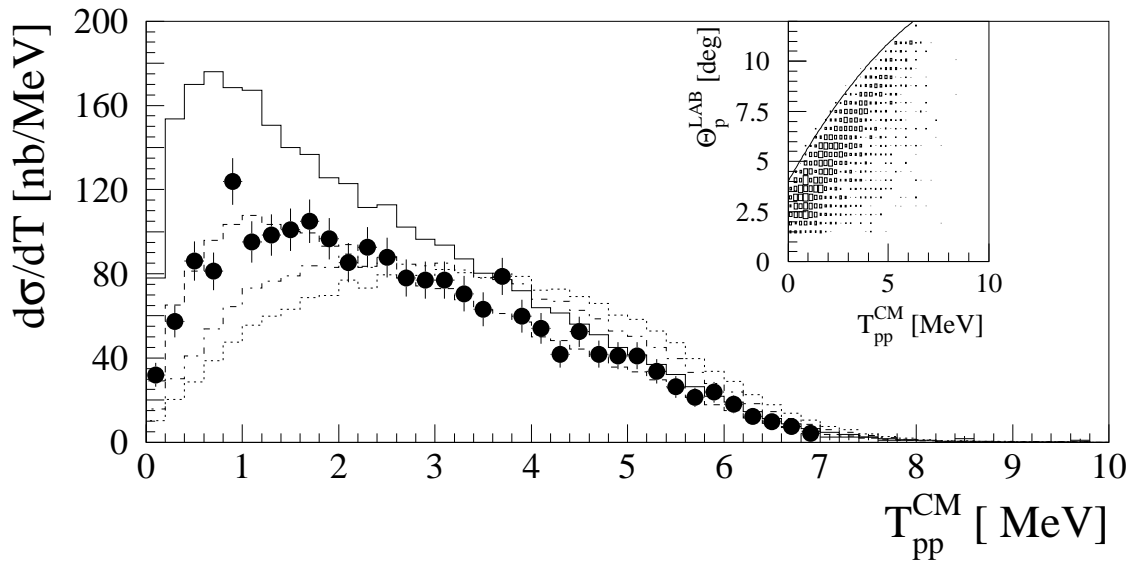,scale=0.8}
\end{center}
\vspace*{-8.2cm}
\caption{\it Proton energy distribution $T_{pp}^{CM}$ in the 2-proton CM
system (solid dots with statistical error bars) in comparison with the 
Monte Carlo simulation including FSI with standard parameters (dashed 
histogram). The extrapolation to full acceptance is shown by the solid 
histogram. The pure phase space distribution is given by the dotted 
histogram, a simulation with FSI and a scattering length $a_0=-1.5\,fm$ 
by the dash-dotted histogram. The insert shows $\Theta_p^{LAB}$, the 
smaller of the two proton angles, vs. $T_{pp}^{CM}$ (see text). The 
solid line denotes the upper boundary in $\Theta_p^{LAB}$ as deduced
from our MC simulation, the lower one coincides with the x-axis showing
clearly the limit in our acceptance slightly below $2^\circ$.} 

\end{figure}

The proton energy distribution $T_{pp}^{CM}$ at 293.5 MeV in the 
center of mass system is shown in fig.~8. That close to threshold proton
energies only range up to roughly 7 MeV with a very distinct maximum 
near 1 MeV. The FSI-based simulation given by the dashed histogram 
reproduces the data very well, whereas a simple phase space distribution
(given by the dotted histogram and normalized to the experimental yield)
fails completely. Also shown is the result of a simulation where final 
state interaction effects have been included but with a much reduced 
scattering length of $a_0$=-1.5 fm as was suggested in \cite{hom2} 
(fig.~8, dash-dotted histogram). When normalizing this distribution to 
our experimental data as was done in fig.~8 very poor agreement is 
observed. If on the other hand one normalizes only to the upper part 
beyond $T_{pp}^{CM}$=3.0 MeV the maximum around 1.0 MeV is totally 
missed and we cannot see how a proper extrapolation can be achieved. 
From the insert showing in a two-dimensional plot $\Theta_p^{LAB}$, the 
smaller of the two proton angles, versus $T_{pp}^{CM}$, it can be seen, 
that the maximum yield found near $T_{pp}^{CM}$=1 MeV is connected 
with one of the two protons emerging towards $\Theta$-values well 
below $5^{\circ}$, a range that is almost completely without detector 
coverage in each of the two internal target experiments. In view of 
this observation we argue that a description of the data and hence an 
extrapolation towards full acceptance can only be performed correctly 
by using the standard values $a_0$=-7.83 fm and $r_0$=2.8 fm as 
scattering length and effective range, respectively. The result is 
given by the solid histogram in fig.~8.
  
One important remark should be added. In the work of Flammang et al. on 
the $pp\to pn\pip$ reaction it was argued that the acceptance
calculation and thus the extraction of a total cross section is 
insensitive to the presence, absence or details of the FSI model
\cite{flam}, since the additional weight factors used for the
extrapolation cancel out. This statement is true to some extent and has 
been checked in the present analysis; the extrapolation to full 
acceptance as given in fig.~8 by the solid line could also be 
obtained from our {\it experimental} points by neglecting in the Monte
Carlo simulation any FSI effects and assuming purely phase space
distributed events. The agreement between experimental and simulated
data in that case, however, would have been of that quality shown in 
fig.~8 by the dotted line. Instead, and we like to stress this as a 
crucial point, our initial distribution, although modified by the 
detector acceptance, not only showed the enhancement due to FSI, but 
this distribution was also well reproduced by our Monte Carlo 
simulation, which we had based on i) our knowledge of detector 
performance and acceptance  
and ii) the use of standard FSI parameters for scattering length and 
effective range. If we would have started from a distribution with 
(almost) no enhancement nothing would have been gained from an 
extrapolation where one tries to adjust the $w_{fsi}$ weights alone.

%\newpage
\vspace{0,5cm}

{\bf C. Angular Distribution}

Acceptance corrected angular distributions for the pion- and relative 
two-proton-momentum vectors are shown in fig.~9 for the measurement at 
T=293.5 MeV. The solid lines are the result of fits in terms of 
Legendre polynomials $P_l$ to the experimental data $W(cos \Theta) 
\propto 1 + \sum_{l}^{}a_l\cdot P_l(cos \Theta)$ up to $l=2$. Only very 
small $a_2$-coefficients were found as expected that close to threshold,
namely $a_2=+0.05 \pm 0.10$ for the top and $a_2=+0.08 \pm 0.11$ for 
the bottom distribution. Similar results were obtained
at the other two bombarding energies. The observed symmetry around 
$cos\,\Theta = 0$ (or the vanishing of the $P_1$-term) which stems from 
the entrance channel consisting of two identical particles was not used 
as a constraint in the fit; instead we consider it as a valuable check 
on the quality of our acceptance corrections.

\begin{figure}[htb]
\vspace*{+0.5cm}
\begin{center}
\epsfig{file=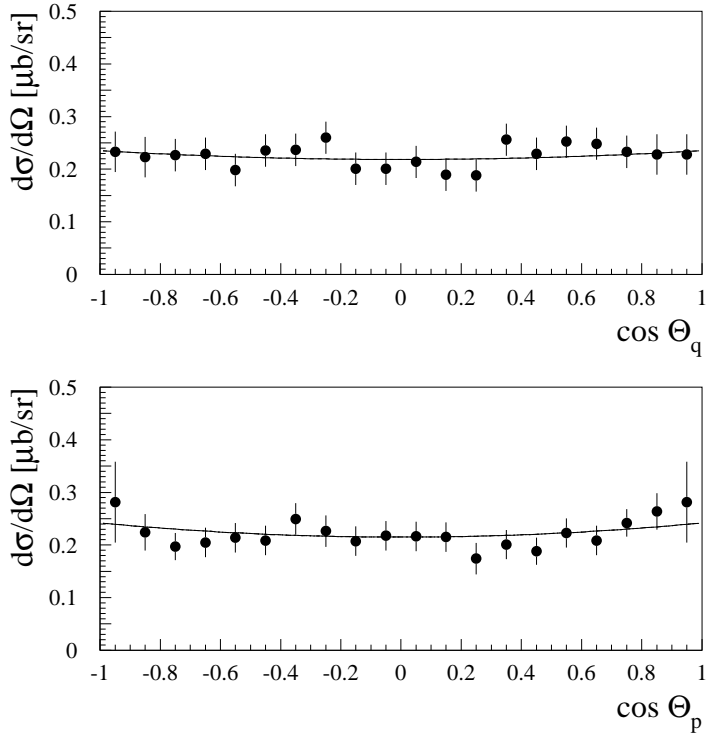,scale=0.5}
\end{center}
\vspace*{-1.5cm}
\caption{\it Acceptance corrected angular distributions in 
$cos\,\Theta _q$ (direction of pion momentum) and $cos\,\Theta _p$ 
(direction of relative two-proton momentum) in the CM system as
obtained at T=293.5 MeV. The solid lines are the result of
Legendre polynomial fits up to $l=2$.}

\end{figure}

The angular distribution obtained for the $pp \to pn\pip$ reaction at 
T=298.1 MeV corresponding to $\eta = 0.19$ was found to be consistent
with isotropy which is in agreement with the published data \cite{flam}.
Sizeable anisotropies were found for the angular distribution of the 
2-particle reaction $pp \to d\pip$ yielding an $a_2$-coefficient of 
order +0.2 and larger with, however, an error of comparable size. This 
is a consequence of the significant acceptance corrections which had to 
be performed to account for those pions which had decayed before 
reaching the detector, as well as those which were too low in momentum 
as being able to reach the second layer of the endcap (see also 
sect.~III.C).

\newpage
%\vspace*{0.5cm}

\begin{large}

{\bf V. Discussion}

\end{large}

The total cross sections listed in table 1 are compared to results
found in the literature \mbox{$\lbrack$ 1-3 $\rbrack$},
\mbox{$\lbrack$ 5-6 $\rbrack$} in fig.~10. In the range  $\eta = 0.2 - 
1.0$ $\sigma$ rises over two orders of magnitude up to values close 
to 100 $\mu$b. The present data exceed the ones from IUCF \cite{hom1}
and CELSIUS \cite{bond} by roughly $50\,\%$. We connect this 
discrepancy with the one basic difference in experimental set-up, namely
theirs being an internal, ours, however, an external target experiment. 
The most critical step in extracting an absolute cross section 
lies in the acceptance correction, which depends on the knowledge 
of i) geometrical coverage of the available phase space, ii) angular 
distribution effects, and iii) deviations from s-wave distributed 
events due to final state effects. Since, most probably,
the geometrical coverage is well accounted for in all three cases, and 
angular distribution effects are negligible, we believe an underestimate
of the strong FSI effects to be responsible for the large
discrepancy. Our reasoning is as follows: The lower the energy above 
threshold, the more important the relative weight of FSI becomes, and,
simultaneously, the more yield gets compressed into the very small angle
region. If on the other hand the minimum angle is only near $4^\circ$,
as is the case in the internal target experiments at IUCF
\cite{hom1} or CELSIUS \cite{bond}, more and more yield is 
missed. To extrapolate into the uncovered region requires a very 
profound knowledge of the original distribution, and most often some 
model dependence will enter the extrapolation procedure. As was 
mentioned before the data given in \cite{hom1} were corrected by using 
non-standard FSI parameters (scattering length $a_0$=-1.5 fm), an
approach which completely failed to reproduce our data. The low-energy 
CELSIUS-data \cite{bond} were obtained by detecting the two decay 
photons instead of the two emerging protons, a method with an 
acceptance which is only weakly dependent on excess energy and is free 
of any threshold effects, but suffers from the very small geometric 
coverage of only $5\,\%$. A rather elaborate acceptance correction has 
been developed for the high-energy CELSIUS data \cite{bilg} where
FSI- and angular distribution effects were considered. As can be seen 
from fig.~10 these data exceed the previous IUCF data \cite{hom1} 
as well although by a smaller margin of some $15\%$.

%\newpage
\begin{figure}[htb]
\vspace*{-1.5cm}
\begin{center}
\epsfig{file=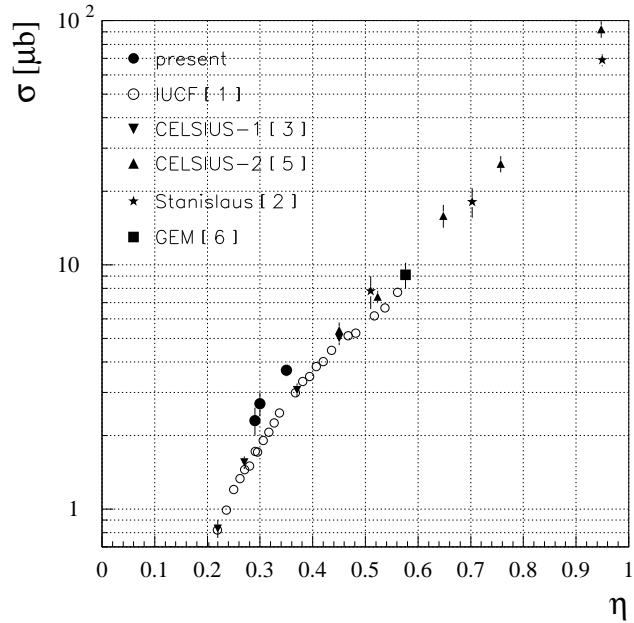,scale=0.5}
\end{center}
\vspace*{-1.5cm}
\caption {\it Summary of cross section data as known today for $pp\to
pp\piz$ up to $\eta=1.0$. The present data given by solid dots clearly 
are higher than the earlier ones from IUCF (open circles, \cite{hom1})
and CELSIUS (solid downward triangles, \cite{bond}).}

\end{figure}

\begin{figure}[htb]
\vspace*{-1.5cm}
\begin{center}
\epsfig{file=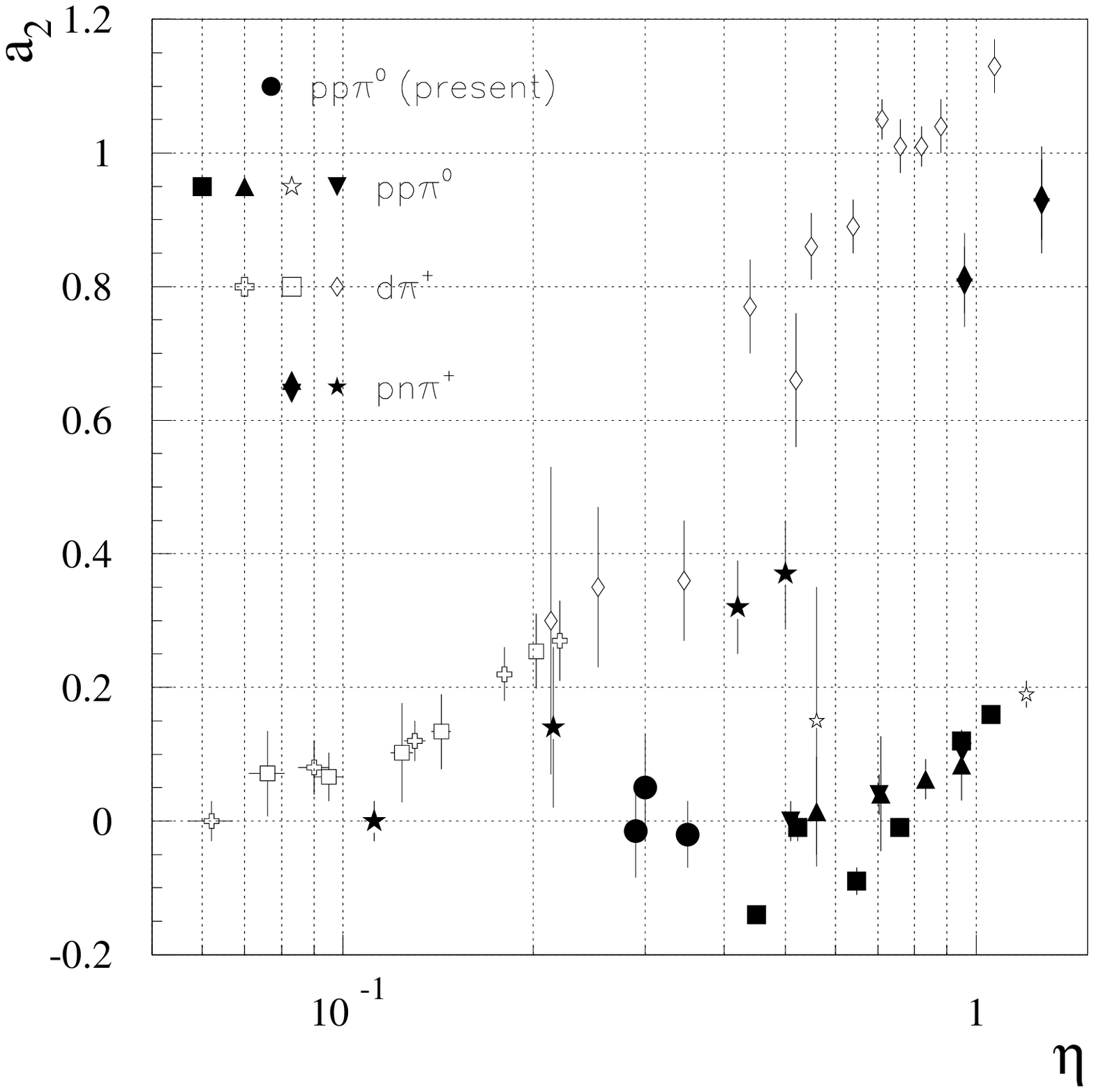,scale=0.5}
\end{center}
\vspace*{-1.0cm}
\caption{\it Comparison of experimental $a_2$-coefficients as obtained
from Legendre polynomial fits to $\pi$-angular distributions in the
near threshold region. $pp\to d\pip$-data are given by open symbols,
the ones for $pp\to pn\pip$ by solid asterisks and diamonds. Solid dots,
squares and triangles as well as open asterisks are used to denote the
data for $pp\to pp\piz$ (see text).}

\end{figure}

A summary of Legendre polynomial expansion coefficients $a_2$ found in
the literature for the pion producing reactions in the near threshold 
region is shown in fig.~11 as a function of $\eta$. Three bands
of data are discernible. The fastest rise with $\eta$ is observed for
the $pp\to d\pip$ reaction (given by open symbols, refs.~$\lbrack 10-11
\rbrack$, \cite{ritc}) where deviations from $a_2 = 0$ are already 
found below $\eta = 0.1$. The positive 
$a_2$-coefficients are an indication for strong p-wave contributions. 
Much less information is available for the other $\pip$-producing 
reaction measured at IUCF and TRIUMF and plotted as asterisks 
\cite{flam} and diamonds \cite{pley}, respectively. Yet a comparable 
behaviour of a fast rise can be observed as well, which, however, only 
starts somewhat later near $\eta = 0.2$. For both these reactions with 
an isoscalar two-nucleon pair in the exit channel the Sp partial wave 
which is not Pauli forbidden contributes almost from threshold. A 
totally different pattern is observed in case of the $pp \to 
pp\piz$-reaction. Near-isotropy is found for almost the whole 
$\eta$-range shown in fig.~11 with $a_2$ staggering between -0.2 and 
+0.2. It is not clear whether d-wave contributions which induce
negative $a_2$ coefficients might be present that close to threshold.
A unique trend (toward positive values) is only observed for $\eta > 
0.9$.

To get a more quantitative estimate on the amount of p-wave 
contributions to the $pp\piz$-reaction at $\eta$ = 0.3 we used the fact
that close to threshold the dynamics for pion production as given by 
the matrix element is almost energy independent. Instead the 
rapid rise in cross 
section is governed by kinematical effects since phase space as well 
as the radial wave functions of the participating particles change 
drastically. In a non-relativistic treatment the available phase space 
volume scales as q$\cdot$p. The wave functions can be approximated by 
their asymptotic form which is valid for small arguments, and from 
this one obtains a factor $q^{l_q}\cdot p^{l_p}$ with $l_q$ ($l_p$) 
denoting the angular momentum of the pion (the NN pair), respectively. 
Hence each of these additional terms has a particular momentum 
dependence as e.g., $\sigma_{Ps}\propto q\cdot p^3$ and  $\sigma_{Pp}
\propto q^3\cdot p^3$ \cite{rose}. To account for these higher order 
terms we introduced into our Monte Carlo code additional weight factors 
$w_{Ps}\,=\,C^2_{Ps}\cdot qp^3$ and $w_{Pp}\,=\,C^2_{Pp}\cdot p^3q^3$ 
with adjustable fit parameters $C^2_{Ps}$ and $C^2_{Pp}$, which were 
added to the dominant weight $w_e\cdot w_{fsi}$ discussed in section 
IVB. Due to the limited statistical accuracy in our data we did not 
attempt a combined fit where the Ps and Pp fractions were varied
simultaneously. Instead we fitted our experimental q- and p-momentum
distributions by a Monte Carlo generated sum of either Ss+Ps or Ss+Pp 
terms. In view of the good overall acceptance of our detector with 
only a small central hole of slightly less than $2^{\circ}$ any 
acceptance dependence of the weight factors, however, was neglected. 
The resulting $\chi^2$-values as obtained for various fractions 
f are shown in table 2 where f is given through $\sigma_{Ps(p)} = 
f\cdot\sigma$ and $\sigma_{Ss} = (1-f)\cdot\sigma$.
The best fit with a reduced 
$\chi^2$/n = 1.24 with n=65 degrees of freedom is found for pure s-wave 
behaviour. Up to $10\%$ of a Pp contribution is allowed by the fit, 
Ps contributions are even weaker. As a result we assign to the 
Ss-strength for the pp$\piz$ reaction at $\eta$=0.3 a fraction 
(1-f)=0.95$\pm$0.05, leaving only a minute contribution of 
0.05$\pm$0.05 for Pp and Ps waves.

\begin{table}

\caption {\it Reduced $\chi^2$ values derived from fitting Monte Carlo 
generated momentum distributions to our experimental data by varying
the amount of Ps or Pp contributions (see text).} 

\vspace{0,5cm}
\begin{center}

\begin{tabular}{|c|cccc|cccc|}

\hline
& & & & & & & & \\
$L_pl_q$ & Ps & & & & Pp & & & \\
& & & & & & & & \\
\hline
& & & & & & & & \\
f & 0 & 0.05 & 0.10 & 0.20 & 0 & 0.05 & 0.10 & 0.20 \\
$\chi^2/n$ & 1.24 & 1.41 & 1.93 & 4.15 & 1.24 & 1.26 & 1.36 & 1.90 \\
& & & & & & & & \\
\hline

\end{tabular}
\end{center}
\end{table}

In fig.~12 we show the presently deduced partial cross sections as a 
function of $\eta$ together with data found in the literature 
\mbox{$\lbrack$ 4-6 $\rbrack$}. Although some discrepancies in 
the data exist, which we attribute to the different methods applied, 
the general trend seems to be pretty much established. Up to $\eta$=0.7
Ss given by the diagonally hatched band is strongest, but gets almost
negligible above $\eta$=0.9. At the same time the Pp strength 
(cross-hatched band) starts to dominate, but saturates near 0.6. Ps
strength (vertically hatched band) seems to be present already below
$\eta$=0.4 and exceeds Ss near $\eta$=0.8. In the crossover region the 
discrepancies are the largest which most probably is due to the 
different methods used. The studies carried out at CELSIUS \cite{bilg} 
and COSY-GEM \cite{gem1} were also performed with unpolarized beam and
target; from these experiments sizeable Pp strength was 
reported well below $\eta$=0.6. A much higher sensitivity can be 
expected when measuring polarization observables as was recently done 
at IUCF \cite{hom2}. In that study $\sigma_{Ps}$ was found to be larger 
than $\sigma_{Pp}$ near $\eta$=0.56.

\begin{figure}[htb]
\vspace*{-1.5cm}
\begin{center}
\epsfig{file=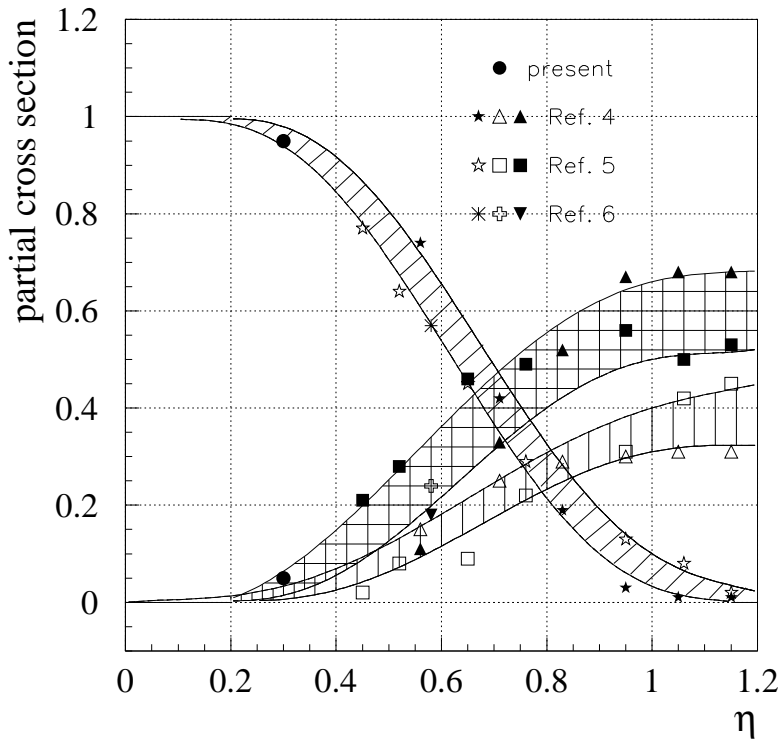,scale=0.6}
\end{center}
\vspace*{-1.5cm}
\caption{\it Summary of partial cross sections $\sigma_{Ll}/\sigma$ 
vs.~$\eta$. The present results are given by the full circles. Solid 
symbols (squares and triangles) which form the cross-hatched band 
denote $\sigma_{Pp}$, open squares and triangles in the vertically 
hatched band represent $\sigma_{Ps}$. $\sigma_{Ss}$ is given by stars 
and asterisks (diagonally hatched band).}

\end{figure}

Up to $\eta = 0.7$ the Ss partial wave dominates the $pp\piz$-reaction. 
As was pointed out in \cite{han1} the cross section in this region 
could only be reproduced satisfactorily when adding an extra heavy 
meson exchange (HME) contribution with a ``free'' strength parameter to 
the coherent sum of direct and rescattering term, both of which make up
roughly $60\,\%$ of the measured yield. Yet at energies above $\eta = 
1$ where even higher partial waves with $l\geq 2$ start to contribute 
significantly and where $\Delta$-isobar contributions have to be 
considered the predictions underestimate the data more and more. In 
view of the present findings of a much larger cross section also at low 
$\eta$ values which would require an even larger strength parameter for 
HME contributions it might be conceivable that something more essential 
is missing in the theoretical description of the near-threshold 
$pp\piz$ reaction. It seems to be worth mentioning that poor agreement 
is also found in recent calculations of polarization observables 
\cite{han3}.

\vspace*{0.5cm}
%\newpage

{\bf VI. Summary}

The $pp \to pp\piz$ reaction has been investigated in a kinematically
complete experiment by detecting the two protons in the large
acceptance spectrometer COSY-TOF set up on an external beam line of
the proton synchrotron COSY. The calibration of the detector and the 
luminosity determination was carried out by analysing the pp-elastic 
scattering reaction recorded simultaneously. Extensive Monte Carlo 
simulations were carried out in order to determine the overall 
acceptance with high precision. Results derived for the concurring
reactions $pp\to d\pip$ and $pp \to pn\pip$ are in very good 
agreement with data from the literature. Total cross sections for the
reaction $pp\to pp\piz$ were obtained at bombarding energies of 292.2, 
293.5 and 298.1 MeV, which exceed previous measurements by roughly 
$50\,\%$. Invariant mass and energy distributions are governed by 
strong final state effects. The importance of detecting protons well 
below $4^{\circ}$ to the beam axis is stressed. The angular 
distributions for the pion and relative two-proton momentum vectors 
were measured for $\eta$-values as low as $\eta=0.3$ and were found 
to be nearly isotropic as expected that close to threshold. From an 
analysis of the corresponding momentum distributions the combined 
fraction for Ps and Pp partial cross sections was estimated to be 
$5\pm5\,\%$.

\vspace*{0.5cm}

{\bf Acknowledgements}

The help of the COSY crew in delivering a low-emittance proton beam is
gratefully acknowledged. The data are based in part on the analysis
work performed by B.~Jakob \cite{jakob}. We like to thank R.~Klein and 
M.~W\"urschig-P\"orsel for their continued assistance in solving
technical problems. Helpful discussions with C.~Hanhart are very much
appreciated. Financial support was granted by the FFE fund
of the Forschungszentrum J\"ulich and by the German BMBF. 

\newpage
%\vspace*{0.5cm}

\end{sloppypar}

\end{document}